\documentclass[preprint,10pt]{elsarticle}
\usepackage[utf8]{inputenc}
\usepackage{amssymb}
\usepackage{amsmath}
\usepackage[cal=boondox]{mathalfa}
\usepackage[mathscr]{eucal}
\usepackage[utf8]{inputenc} 
\usepackage[T1]{fontenc}    
\usepackage{hyperref}       
\usepackage{url}            
\usepackage{booktabs}       
\usepackage{amsfonts}       
\usepackage{nicefrac}       
\usepackage{microtype}      
\usepackage{lipsum}
\usepackage{afterpage}
\usepackage{graphicx,xcolor}
\usepackage{float}
\usepackage[ruled]{algorithm2e}
\usepackage{breqn}
\usepackage{afterpage}
\usepackage{array}
\usepackage{makecell}

\journal{Physica B}

\begin{document}
\begin{frontmatter}

\title{Do interactions among unequal agents undermine those of low status?}

\author[LISC,LAPSCO]{Guillaume Deffuant}
\ead{guillaume.deffuant@inrae.fr}
\author[LISC]{Thibaut Roubin}

\address[LISC]{Université Clermont Auvergne, INRAE, UR LISC, Aubi\`ere, France}
\address[LAPSCO]{Université Clermont Auvergne, LAPSCO, Clermont-Ferrand, France.}

\begin{abstract}
We consider a recent model in which agents hold opinions about each other and influence each other's opinions during random pair interactions. When the opinions are initially close, on the short term, all the opinions tend to increase over time. On the contrary, when the opinions are initially very unequal, the opinions about agents of high status increase, but the opinions about agents of low status tend to stagnate without gossip and to decrease with gossip. We derive a moment approximation of the average opinion changes that explains these observations.
\end{abstract}

\begin{highlights}
\item We aim at explaining patterns generated by an already published model in which agents discuss and modify their opinions about each other and about themselves;
\item We derive a mathematical approximation of the average change of the opinions for a random interaction and we assess its accuracy;
\item The opinions about an agent tend to evolve similarly and their common evolution is determined by a weighted sum of the increasing tendency of the self-opinion and the decreasing tendency of the opinions about others;
\item The analysis of this weighted sum shows that, the lower the status of the agent, the more the opinions about the agent tend to decrease, especially when gossip is activated, while the opinions about agents of high status tend to increase;
\item This analysis provides new explanations to the patterns generated by the model.
\end{highlights}


\end{frontmatter}

\section{Introduction}
Most opinion dynamics models consider opinions\footnote{When using "opinions", we conform to the usage in the research community but we think that for our model, "attitudes" would be more appropriate.} about objects, like commercial products, or  about actions on the world, like political options (\cite{French1956,Galam2002,Deffuant2000,Deffuant2006a,Hegselmann2002,Flache2011}, for a recent review see: \cite{Flache2017}).  It is generally assumed that when agents discuss about an object or an action, they influence each other's opinions. These interactions can therefore determine commercial or political successes or failures. 

In these approaches, apart from a few exceptions \cite{Bagnoli2007,Carletti2011}, opinions about the agents themselves are not considered as deserving any specific attention. However, the opinions about agents determine the social network of positive or negative connections, hence in some respect the social structure.  Moreover, it is generally recognised that this social structure has a strong influence on the agents' opinions. This suggests that opinions about agents do matter.

This paper precisely focuses on the dynamics of opinions about agents. It builds on previous research \cite{Deffuant2018,Deffuant2013} on models of agents that hold an opinion (a real number between -1 and +1) about all the others and themselves. The model dynamics repeats encounters of two randomly chosen agents influencing their self-opinions and their opinions about each other. Moreover, if gossip is activated, both agents influence their opinions about some other randomly chosen agents. The influence is attractive and the agents are more influenced by the ones that they hold in high esteem. Importantly, agents do not have a transparent access to the opinions of others; they constantly make errors of interpretation which are modelled by a random noise.

A strong assumption of the approach is that the self-opinion of agent ego is shaped by its perception of the opinions of others about ego. Therefore, on average, ego's self-opinion measures how ego feels perceived by others. As stressed in \cite{Huet2020}, this is in line with the hypothesis considering self-esteem as a sociometer \cite{Leary2005}.  

The approach also postulates a pivotal role of self-opinions in the influence function. Moreover, the initial version of the model \cite{Deffuant2013} includes an additional dynamics, called vanity, in which the self-opinion plays a major role. When combining the attractive dynamics and vanity, a significant positive bias on self-opinions emerges: ego's self-opinion is significantly higher than the average of the opinions of others about ego (see details in \cite{Deffuant2013}).  Further investigations show that even without vanity, the model generates intriguing patterns and the self-opinions are also slightly higher than the opinions about the agent \cite{Huet2017,Deffuant2018}. 

The main contribution of this paper is an analytical approximation of the average (first moment) evolution of the opinions in the model and the average evolution of their products (second moment). This moment approximation is inspired by a general approach already applied on other agent based models \cite{Law2000}. This moment approximation confirms the hypothesis (formulated in \cite{Deffuant2018}) that the evolution of the opinions about an agent is determined by a combination of positive effects on the agent self-opinion and negative effects on the opinions of others about this agent. Moreover, the moment approximation explains why the opinions about the agents of low status tend to stagnate or to decrease (especially when there is gossip) while the opinions about agents of high status tend to increase. Finally, these results explain the patterns described in \cite{Deffuant2018}.

The following section describes the model and presents the patterns in more details. Section 3 is devoted to the moment approximation. Section 4 analyses the accuracy of the approximation and studies the effect of initial opinion inequalities in the case of a group of 10 agents. The last section is devoted to a discussion about the results and their possible connections with some research in social-psychology. 

\section{Model and patterns}
This section first presents the model in details and the main patterns emerging from its dynamics that drew attention in previous research. Then, it recalls their hypothetical explanation, proposed in \cite{Deffuant2018}.

\subsection{The model}
\label{sec:model}
The model is the same as in \cite{Deffuant2018}. We present it here with slightly different notations. It includes $N_a$ agents. Each agent $i \in \{1,\dots, N_a\}$  has an opinion $a_{ij}$ about each agent $j \in \{1,\dots, N_a\}$ including themselves. The opinions are real values between -1 and +1. In \cite{Deffuant2018}, at the initialisation, all opinions are set to 0: agents have a neutral opinion about themselves and all the others at the beginning of the simulations. In this paper, we shall also consider specific initial values of the opinions expressing different levels of initial perceived inequalities.

Graphically, the agents' opinions can be represented as a matrix (see examples on Figure \ref{fig:noGroupNoGossipPatterns}) in which row $i$, with $1 \leq i \leq N_a$, represents the array of $N_a$ opinions of agent $i$ about the agents $j$. Column $j$, with $1 \leq j \leq N_a$, represents the opinions all agents $i$ about $j$. Positive opinions are represented with red shades and negative opinions with blue shades. Lighter shades are used for opinions of weak intensity (close to 0). 




The dynamics consists in repeating:
\begin{itemize}
    \item choose randomly two distinct agents $i$ and $j$;
    \item $i$ and $j$ interact: $j$ influences $i$'s opinions and $i$ influences $j$'s opinions.
\end{itemize}
  
In this interaction, $a_{ii}(t)$, $i$'s self-opinion, is influenced by $a_{ji}(t)$, the opinion of $j$ about $i$. As a result of this influence, $a_{ii}(t)$ gets closer to a noisy evaluation of $a_{ji}(t)$. The modification of $a_{ii}(t)$, denoted by $\Delta a_{ii}(t)$, is ruled by the following equation, in which $\theta_{ii}(t)$ designates a number that is uniformly drawn between $-\delta$ and $\delta$ ($\delta$ being a parameter of the model): 

\begin{align} \label{eq:deltas1}
\Delta a_{ii}(t) &=  h_{ij}(t) (a_{ji}(t)-a_{ii}(t) + \theta_{ii}(t)),
\end{align}
Similarly, the change of opinion $a_{ji}(t)$, is:
\begin{align} \label{eq:deltas3}
\Delta a_{ji}(t) &=  h_{ji}(t) \left(a_{ii}(t)-a_{ji}(t) + \theta_{ji}(t)\right).
\end{align}
where  $\theta_{ji}(t)$, is a uniformly drawn number between $-\delta$ and $\delta$. The function of influence $ h_{ij}(t)$ is given by equation \ref{eq:credibility}, expressing that the more $i$ perceives $j$ as superior, the more $j$ is influential on $i$.

\begin{equation}\label{eq:credibility}
 h_{ij}(t) = H(a_{ii}(t)- a_{ij}(t)) = \frac{1}{1 + \exp{\left( \frac{ a_{ii}(t) - a_{ij}(t) }{\sigma}\right)} }\,. 
 \end{equation}In this model, self-opinions measure how well agents think they are perceived by others, with a stronger weight attributed to agents perceived as superior. As stressed in the introduction, this is in line with the hypothesis considering self-opinion as a sociometer \cite{Leary2005}.  



When activating gossip, agents $j$ and $i$ influence their opinions about $k$ agents $g_p$ , $p \in \{1,\dots,k\}$ drawn at random such that $g_p \neq i$ and  $g_p \neq j$. The changes of the opinion of $i$ about agents $g_p$ are:
\begin{align} \label{eq:deltaGossip}
    \Delta a_{ig_p}(t) &= h_{ij}(t) (a_{jg_p}(t)-a_{ig_p}(t) + \theta_{ig_p}(t)), \mbox{ for } p \in \{1,\dots,k\},
\end{align}
where $\theta_{ig_p}(t)$ is a uniformly drawn number between $-\delta$ and $\delta$. The changes of the opinion of $j$ about these agents follow the same equations where $j$ and $i$ are inverted.

Overall, after the encounter between $i$ and $j$, the opinions about $i$ change as follows:
\begin{align}
\label{eq:modif}
    a_{ii}(t+1) &= a_{ii}(t) + \Delta a_{ii}(t),\\
    a_{ji}(t+1) &= a_{ji}(t) + \Delta a_{ji}(t).
\end{align}
The opinions about $j$ change similarly (inverting $j$ and $i$ in the equations). 
If there is gossip ($k > 0$), $k$ agents $g_p$ are randomly chosen with $p \in \{1,\dots,k\}$, $g_p \neq i$ and  $g_p \neq j$, and the opinions about $g_p$, for $p \in \{1,\dots,k\}$ change as follows:
\begin{align}
\label{eq:modifGossip}
        a_{ig_p}(t+1) &= a_{ig_p}(t) + \Delta a_{ig_p}(t),\\        a_{jg_p}(t+1) &= a_{jg_p}(t) + \Delta a_{jg_p}(t).
\end{align}

The opinions are updated synchronously: at each encounter all the changes of opinions (e.g. equations \ref{eq:deltas1}, \ref{eq:deltas3} and \ref{eq:deltaGossip}) are first computed and then the opinions are modified simultaneously (e.g. equation \ref{eq:modif}, \ref{eq:modifGossip}).



Overall, the following parameters tune the dynamics:
\begin{itemize}
\item $\sigma$ defines the shape of the influence function $h_{ij}$; if $\sigma$ is very small, the function is very tilted, meaning that agents are subject to high influence from the ones that they evaluate better than themselves and they almost completely disregard the opinions of the ones considered lower.   
\item $\delta$ represents the amplitude of the uniformly distributed errors that perturb the evaluation of others' expressed opinions.  This noise stands for the inability of an agent to directly access the opinion of others. Without it, from all opinions at zero, there would be no opinion change at all.
\item $k$ is the number of agents subject of gossip in each pair interaction (hence if $k = 0$, there is no gossip).
\end{itemize}

\subsection{The main patterns}
\label{sec:patterns}



Figure \ref{fig:noGroupNoGossipPatterns} illustrates the main patterns of evolution of the opinions reported in \cite{Deffuant2018}.

Panels (a) and (c) illustrate the pattern obtained without gossip ($k = 0$). Panel (a) shows a typical opinion matrix after a large number of iterations. In each matrix column the opinions are close and the differences between the matrix columns are stronger than the differences of opinions within each column. This is explained by the attractive dynamics which tends to align the opinions about a given individual. Note that most of the columns are red, indicating that the opinions about most agents are positive. On panel (c) the red curve shows the evolution of the average opinion.  The blue curves are the evolution of the agents' reputations (the average opinion about an agent). Starting from 0, the average opinion increases and then fluctuates around a significantly positive value (close to 0.5). As already noticed in \cite{Deffuant2018}, this pattern is surprising because, at a first glance, the equations do not privilege changing opinions upward and the noise is symmetric around 0. 
 
Panels (b) and (d), illustrate the pattern taking place when gossip is activated (in this case, $k = 5$). The matrix of opinions after a large number of interactions (panel b) shows  numerous blue columns. On panel (d), the evolution of the average opinion (red curve) remains negative with significant fluctuations while the reputations (blue curves) are more dispersed than without gossip, with a larger density in the low part of the opinion axis. 
 

\afterpage{%
\begin{figure}[!h]
\centering
   \begin{tabular}{ccc}
   No Gossip ($k = 0$) &\hspace{0.5 cm} & Gossip ($k = 5$)\\
   \hspace{0.7 cm} \includegraphics[width= 4.5 cm]{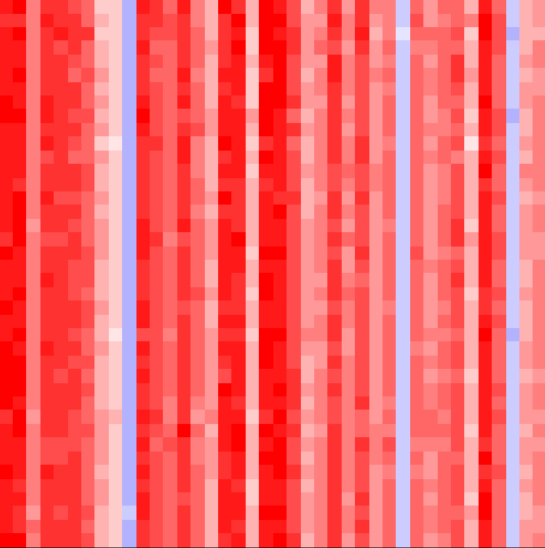} & \hspace{0.5 cm} &
  \hspace{0.5 cm}   \includegraphics[width= 4.5 cm]{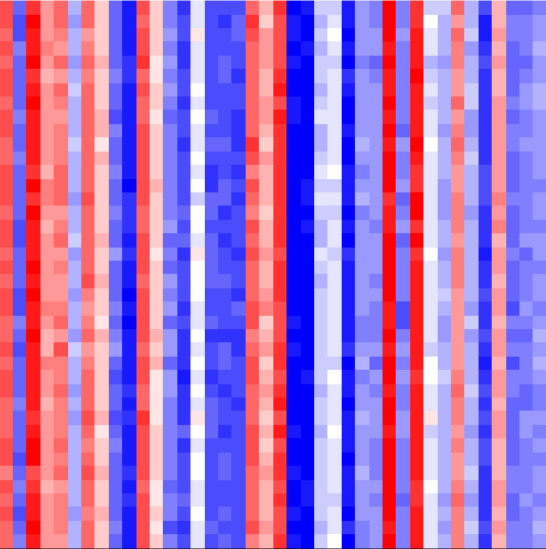} \\
		(a) & \hspace{0.5 cm} & (b) \\
    \includegraphics[width= 5.5 cm]{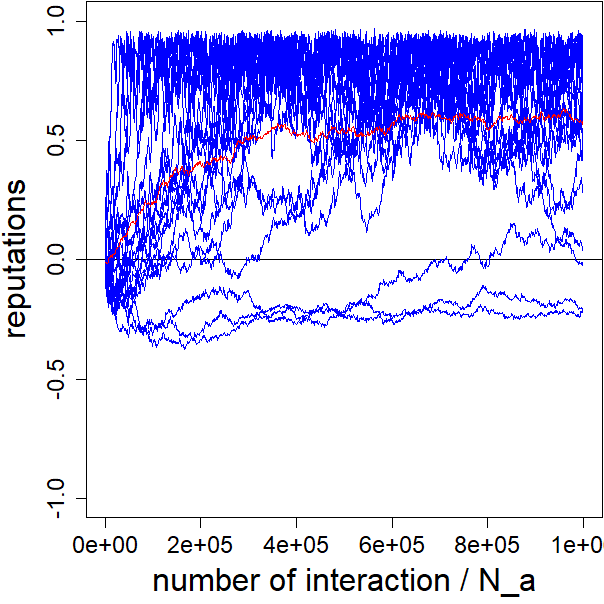} & \hspace{0.5 cm} &
    \includegraphics[width= 5.5 cm]{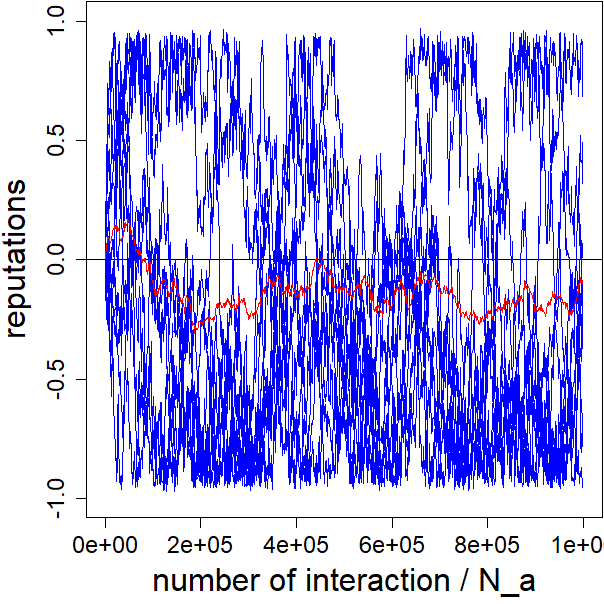} \\
		(c) & \hspace{0.5 cm} & (d)
    \end{tabular}
   \caption{Typical patterns, with $\delta = 0.1$ (noise), $\sigma = 0.3$ (influence function parameter) and $N_a =40$ agents. Panels (a) and (b) show the matrix of opinions after $1$ million $\times N_a$ pair interactions. Panels (c) and (d) show the evolution of the average opinion (in red) and the evolution of the agent reputations (in blue, the reputation of agent $i$ being the average of the opinions about $i$). }
	\label{fig:noGroupNoGossipPatterns}
\end{figure}
\clearpage
}

\subsection{Hypothetical explanation of the patterns.}
In \cite{Deffuant2018}, the patterns are related to two biases which are observed in a simplified setting where only one opinion varies, between two interacting agents:
\begin{itemize}
    \item when the self-opinion of ego varies, it is on average slightly higher than the opinion of alter about ego. There is a positive bias on the self-opinion. 
    \item when the opinion of ego about alter varies, symmetrically, it is on average slightly lower than alter's self-opinion. There is a negative bias on the opinion about others.
\end{itemize}
In the following, paragraph \ref{sec:biases} describes this setting in more details and derives mathematical expressions of the biases.

The authors of \cite{Deffuant2018} hypothesise that similar biases are present when all opinions vary and more than two agents interact. Moreover, they suggest that the drift to positive or negative opinions is due to the dominance of one bias on the other:
\begin{itemize}
    \item Without gossip, the positive bias on self-opinion dominates the negative bias on the opinions about others, which explains why the positive drift arises;
    \item Gossip increases the noise on the opinions about others, which increases the negative bias on opinion about others, leading to its possible domination over the positive bias on self-opinion.
\end{itemize}

This hypothesis is indirectly supported by experiments involving several agents but only the opinions about one specific agent vary. These experiments measure the average evolution of the opinions over a large number of simulations and their results are compatible with the hypothesis. However, these explanations remain very general and qualitative. 

We now derive a moment approximation of the evolution of the opinions, in order to formally define the biases and to determine precisely their connection to the patterns.

\section{Moment approximation.}
We first derive the moment approximation in the case (already presented in \cite{Deffuant2018}) of only one opinion varying between two interacting agents. Then, we extend the approach to the general case of all varying opinions both with or without gossip. Finally, we introduce the equilibrium opinion that determines the effect of second order shared by all the opinions about an agent. 
 
\subsection{Single opinion varying between two interacting agents.}
\label{sec:biases}
We assume that only two agents, $1$ and $2$, interact and firstly only the self-opinion $a_{11}(t)$ is varying, starting from $a_{11}(0) = a$. The other opinions are fixed: for any value of $t$, $a_{12}(t) = b$ (the opinion of $1$ about $2$), $a_{21}(t) = a$ (the opinion of $2$ about $1$), $a_{22}(t) = b$ (self-opinion of $2$). Moreover, for any value of $i$ and $j$, we define $x_{ij}(t)$ as the opinion offset from $t = 0$:
\begin{align}
    x_{ij}(t) = a_{ij}(t) - a_{ij}(0).
\end{align}

At the first step, $1$ perceives $a_{21}(1) $ as $a + \theta(1)$, $\theta(1)$ being drawn from the uniform distribution between $-\delta$ and $\delta$. Applying the interaction rule:
\begin{align}
    x_{11}(1) &=  h_{12}(0)(a + \theta(1) - a) \\
                &=  h \theta(1),
\end{align}
where $h = h_{12}(0) = H(a - b)$. For any expression $y$, let $\overline{y}$ be the average of expression $y$ over all possible values of the noise. Then, $\overline{x_{11}}(1)$ , the average of $x_{11}(1)$ over all possible draws of $\theta(1)$, is:
\begin{align}
    \overline{x_{11}}(1) =  h \frac{\int_{-\delta}^{+\delta}{\theta}d\theta}{2\delta} = 0.
\end{align}
At the second step, applying the interaction rule again gives:
\begin{align}
    x_{11}(2) &= x_{11}(1) + h_{12}(1) (a + \theta(2) - a - x_{11}(1)).
\end{align}

Assuming that $\theta(1)$ is small, approximating the influence function at the first order gives:
\begin{align}
    h_{12}(1) &= H(a + h\theta(1) - b)\\
    & \approx h + h' h\theta(1),
\end{align}
where $h' = H'(a-b)$.
We get:
\begin{align}
     x_{11}(2) &\approx x_{11}(1) + (h + h'h\theta(1))(\theta(2) - h\theta(1)),\\
     & \approx (1-h)h\theta(1) + h (\theta(2) - h\theta(1)) + h'h \theta(1)\theta(2) - h'h^2\theta^2(1).
\end{align}
Since $\overline{x_{11}}(1) = 0$, $\overline{\theta}(2) = 0$ and $\overline{\theta(1)\theta(2)} = 0$, we have:
\begin{align}
   \overline{x_{11}}(2)
    &= - \frac{h' h^2\int_{-\delta}^{+\delta}\theta^2 d\theta}{2 \delta},\\
    & = -\frac{h' h^2 \delta^2}{3}, \label{eq:bias2st}
\end{align}
Formula \ref{eq:bias2st} applies to any function $h$. Therefore, at the second iteration, the average of $x_{11}(2)$ is positive as soon as function $H(a-b)$ is decreasing when $a-b$ is increasing. It is called the positive bias on self-opinions in \cite{Deffuant2018}. 

With the choice of $H$ specified by equation \ref{eq:credibility}, we have $h' = \frac{-h (1-h)}{\sigma}$, hence:
\begin{align}
\label{eq:posBiast2}
    \overline{x_{11}}(2) & = \frac{h^3(1-h)\delta^2}{3 \sigma}.
\end{align}


For any number of iterations $t \geq 2$ it can be shown that:
\begin{dmath}
  \overline{x_{11}}(t)= \overline{x_{11}}(2) \left( \frac{1 - (1-h)^{t-1}}{h} + \frac{(1-h)^2}{h^2}\left(\frac{1 - (1-h)^{2(t-2)}}{2-h}- (1-h)^{t-2}\left(1 - (1-h)^{t-2}\right) \right)\right).
  \label{eq:posBiast}
\end{dmath}
Therefore, for an infinite number of iterations:
\begin{align} \label{eq:posBiasinf}
    \overline{x_{11}}(\infty) & = \frac{-h'\delta^2}{3(2-h)},\\
   &= \frac{h (1-h)\delta^2}{3 \sigma (2-h)}.
\end{align}

Assuming that: $H(b, a) = 1 - H(a, b) = 1 - h$, like for our choice of $H$ (specified by equation \ref{eq:credibility}), it can easily be seen that $ \overline{x_{21}}(2)$ is obtained by replacing $h$ by $1-h$ in the expression of $ \overline{x_{11}}(2)$:
\begin{align}
   \overline{x_{21}}(2) & = \frac{h'(1- h)^2 \delta^2}{3}. 
\end{align}

 If $H(a-b)$ is decreasing then $h'$ is negative and $\overline{x_{21}}(2)$ is negative. With $H$ defined by equation \ref{eq:credibility}, we have:
\begin{align}
\label{eq:negBiast2}
    \overline{x_{21}}(2) \approx - \frac{h (1-h)^3 \delta^2}{3 \sigma}.
\end{align}

Finally, $\overline{x_{21}}(t)$ and $\overline{x_{21}}(\infty)$ are also obtained by replacing $h$ by $1-h$ in the expression of $\overline{x_{11}}(t)$ and $\overline{x_{11}}(\infty)$ (equations \ref{eq:posBiast} and \ref{eq:posBiasinf}) and multiplying them by -1.

\subsection{Evolution of average opinions without gossip}

Now, we consider a set of $N_a$ agents interacting as specified in section \ref{sec:model}. First, we average over the noise in the interactions defined by the sequence of randomly chosen couples $s_t = \{(i_1,j_1),\dots,(i_t,j_t)\}$. Then we average over all possible sequences $s_t$ of randomly chosen couples. 

For $(i,j) \in \{1,\dots,N_a\}^2$, let $a_{ij}(s_t)$ be the opinion of agent $i$ about agent $j$ after the sequence $s_t$ of interactions and  $x_{ij}(s_t)$ be the opinion offset from $t=0$:
\begin{align}
    x_{ij}(s_t) = a_{ij}(s_t) - a_{ij}(0).
\end{align}
For any variable $y(s_t)$, let $\overline{y}(s_t)$ be the average of $y(s_t)$ over the noise during the interactions defined by the sequence of couples $s_t$, and let:
\begin{align}
    h_{ij}(s_t) &= H(a_{ii}(s_t)-a_{ij}(s_t));\\
    \overline{h_{ij}}(s_t) &= H(\overline{a_{ii}}(s_t)-\overline{a_{ij}}(s_t));\\
    \overline{h'_{ij}}(s_t) &= H'(\overline{a_{ii}}(s_t)-\overline{a_{ij}}(s_t)).
\end{align}
We approximate $h_{ij}(s_t)$ at the first order around $\overline{h_{ij}}(s_t)$:
\begin{align}
    h_{ij}(s_t) & \approx \overline{h_{ij}}(s_t) + \overline{h'_{ij}}(s_t)(x_{ii}(s_t) - x_{ij}(s_t)- \overline{z_{ij}}(s_t)),
\end{align}
where :
\begin{align}
    \overline{z_{ij}}(s_t) = \overline{x_{ii}}(s_t) - \overline{x_{ij}}(s_t).
\end{align}

 

 For $(i, j) = (i_{t+1},j_{t+1})$ or $(i,j) = (j_{t+1}, i_{t+1})$, applying the rule of opinion change, we get:
\begin{dmath} \label{eq:xiist}
     x_{ii}(s_{t+1}) = x_{ii}(s_t) +\\ \left(\overline{h_{ij}}(s_t) +  \overline{h'_{ij}}(s_t)\left(x_{ii}(s_t) - x_{ij}(s_t) - \overline{z_{ij}}(s_t)\right)\right)\left(x_{ji}(s_t) + \theta_{ii}(t) - x_{ii}(s_t)\right).
\end{dmath}
For sake of simplicity, we assume that all the opinions about any agent $i$ are the same at $t = 0$: $a_{ii}(0) - a_{ji}(0) = 0$ for all couples $(i,j) \in \{1,...,N_a\}^2$. Indeed, when $a_{pi}(0) - a_{ji}(0) \neq 0$ the interactions tend rapidly to drive all the opinions about an agent to a very close value, hence this assumption is not restrictive.

Moreover, it can easily be checked that the average product of opinions about two different agents is always zero, hence $\overline{x_{ii}(s_t).x_{jj}(s_t)} = 0$, $\overline{x_{ij}(s_t).x_{ii}(s_t)} = 0$ and $\overline{x_{ij}(s_t).x_{ji}(s_t)} = 0$.
 Therefore, since $\overline{\theta_{ii}}(t) = 0$, averaging equation \ref{eq:xiist} and neglecting the terms of order higher than 2 yields:
 
 \begin{dmath}
 \label{eq:avm}
      \overline{x_{ii}}(s_{t+1}) = \overline{x_{ii}}(s_t) + \widehat{h_{ij}}(s_t)\left( \overline{x_{ji}}(s_t) -  \overline{x_{ii}}(s_t) \right) \\ +  \overline{h'_{ij}}(s_t) \left(\overline{x_{ji}(s_t).x_{ii}(s_t)} -  \overline{x^2_{ii}}(s_t)\right),
\end{dmath}
with $\widehat{h_{ij}}(s_t) = \overline{h_{ij}}(s_t) -  \overline{h'_{ij}}(s_t)\overline{z_{ij}}(s_t)$.


Applying the same approach to $x_{ji}(s_{t+1})$, we get:
\begin{dmath}
\label{eq:avy}
      \overline{x_{ji}}(s_{t+1}) = \overline{x_{ji}}(s_t) + \widehat{h_{ji}}(s_t)\left( \overline{x_{ii}}(s_t) -  \overline{x_{ji}}(s_t) \right)  \\ + \overline{h'_{ji}}(s_t) \left(   \overline{x^2_{ji}}(s_t) - \overline{x_{ji}(s_t).x_{ii}(s_t)}\right).
 \end{dmath}
 
For $(i, j) = (i_{t+1},j_{t+1})$ or  $(j, i) = (i_{t+1},j_{t+1})$, we can similarly derive the value of the second moment $\overline{x_{ii}^2}(s_{t+1})$. Neglecting the terms of degree higher than 2, we get: 
\begin{dmath}
 \label{eq:avaiisq}
      \overline{x^2_{ii}}(s_{t+1}) =\overline{x^2_{ii}}(s_{t}) + 2 \widehat{h_{ij}}(s_t)\left(\overline{x_{ii}(s_{t})x_{ji}(s_t)} - \overline{x^2_{ii}}(s_{t})\right) +  \widehat{h_{ij}}(s_t)^2 \left(\overline{x^2_{ji}}(s_{t}) + \overline{x^2_{ii}}(s_{t}) - 2\overline{x_{ii}(s_{t})x_{ji}(s_t)}\right)+ \overline{h_{ij}}^2(s_t)\frac{\delta^2}{3}.
\end{dmath} 

Similarly, for $\overline{x_{ji}^2}(s_{t+1})$:
\begin{dmath}
 \label{eq:avajisq}
      \overline{x^2_{ji}}(s_{t+1}) = \overline{x^2_{ji}}(s_{t}) + 2 \widehat{h_{ji}}(s_t)\left(\overline{x_{ii}(s_{t})x_{ji}(s_t)} - \overline{x^2_{ji}}(s_{t})\right) +  \widehat{h_{ji}}(s_t)^2 \left(\overline{x^2_{ji}}(s_{t}) + \overline{x^2_{ii}}(s_{t}) - 2\overline{x_{ii}(s_{t})x_{ji}(s_t)}\right)+ \overline{h_{ji}}^2(s_t)\frac{\delta^2}{3}.
\end{dmath} 

In both cases, the last term of the equation uses the result obtained in section \ref{sec:biases}, for any interaction noise $\theta(t)$:
\begin{align}
    \overline{\theta^2}(t) = \frac{\delta^2}{3}. 
\end{align}

Using a similar approach, we compute the expressions of $\overline{x_{ii}(s_{t+1}).x_{ji}(s_{t+1})}$ and $\overline{x_{pi}(s_{t+1}).x_{ji}(s_{t+1})}$, for $(i,j, p) \in \{1,\dots, N_a\}^2$ (see appendix section \ref{sec:nogos}). 

Now, we average the previous equations over all possible sequences of interactions $s_t$. For any expression $\overline{y}(s_t)$, let $\overline{y}(t)$ be the average of $\overline{y}(s_t)$ over all interaction sequences $s_t$.
Drawing couple $(i,j)$ or couple $(j,i)$ at $t$ has the probability $\frac{2}{N_a(N_a -1)}$, hence averaging equation \ref{eq:avm} over all possible sequences $s_t$ yields:
 \begin{dmath}
 \label{eq:avavm}
      \overline{x_{ii}}(t+1) =  \overline{x_{ii}}(t) + \frac{2}{N_c} \sum_{j \neq i}\left( \widehat{h_{ij}}(t)\left(\overline{x_{ji}}(t) - \overline{x_{ii}}(t)\right) + \overline{h'_{ij}}(t)\left(\overline{x_{ii}(t).x_{ji}(t)} - \overline{x_{ii}^2}(t) \right)\right),
\end{dmath}
with:
\begin{align}
    N_c &= N_a(N_a-1),\\
    \widehat{h_{ij}}(t) &= \overline{h_{ij}}(t) - \overline{h'_{ij}}(t)\overline{z_{ij}}(t),\\
    \overline{z_{ij}}(t) &= \overline{x_{ii}}(t) - \overline{x_{ij}}(t),\\
    \overline{h_{ij}}(t) &= H(\overline{a_{ii}}(t) - \overline{a_{ij}}(t)),\\
    \overline{h'_{ij}}(t) &= H'(\overline{a_{ii}}(t) - \overline{a_{ij}}(t)).
\end{align}
Similarly, averaging equation \ref{eq:avy} over all possible sequences $s_t$, yields:
 \begin{dmath}
 \label{eq:avavy}
       \overline{x_{ji}}(t+1) =\overline{x_{ji}}(t) + \frac{2}{N_c}\left( \widehat{h_{ji}}(t)\left(\overline{x_{ii}}(t) - \overline{x_{ji}}(t)\right) \\ + \overline{h'_{ji}}(t)\left(\overline{x_{ji}^2}(t) -  \overline{x_{ii}(t).x_{ji}(t)}\right)\right).
\end{dmath}

Moreover, we derive the equations of the second moments $\overline{x_{ii}^2}(t+1)$, $\overline{x^2_{ij}}(t+1)$, $\overline{x_{ii}(t+1).x_{ji}(t+1)}$ and $\overline{x_{pi}(t+1).x_{ji}(t+1)}$ for $(i,j,p) \in \{1,\dots,N_a\}^2$ (see appendix, section \ref{sec:nogosav}). Then, with the initial values of these variables at $t=0$, we can compute the values of $\overline{x_{ii}}(t+1)$ and $\overline{x_{ij}}(t+1)$ for $(i,j) \in \{1,\dots,N_a\}^2$ at any time step $t$ by induction.

For $Na = 2$, we could derive simple direct expressions of $ \overline{x_{ii}}(t)$ and $\overline{x_{ji}}(t)$ (not reported in this paper) but for $N_a > 2$ we only get the values by iterating the formulas until reaching $t$. 
\subsection{Evolution of the average opinions about an agent when gossip is activated.}

Now, in the sequence defining the interactions, to each pair $(i_t, j_t)$ we add a set $(g_{1_t},\dots,g_{k_t})$ of $k$ elements of $\{1,\dots,N_a\}$ distinct from $i_t$ and $j_t$, about which $j_t$ and $i_t$ gossip.
\begin{itemize}
    \item If $(i, j) = (i_{t+1},j_{t+1})$ or $(j, i) = (i_{t+1},j_{t+1})$, the equations of $\overline{x_{ii}}(s_{t+1})$ and $\overline{x_{ji}}(s_{t+1})$ are the same as in the previous paragraph.
    \item Moreover, for any $g \in \{g_{1_{t+1}},\dots,g_{k_{t+1}})$, we have:
\begin{dmath}
       \overline{x_{ig}}(s_{t+1}) = \overline{x_{ig}}(s_{t}) + \widehat{h_{ij}}(s_t)\left( \overline{x_{jg}}(s_{t}) -  \overline{x_{ig}}(s_{t})\right).
\end{dmath}
\end{itemize}

The equations specifying $\overline{x_{ii}^2}(s_{t+1})$, $\overline{x_{ji}^2}(s_{t+1})$ and the other second order moments are specified in the appendix (section \ref{sec:gosav}).                     

Now, we derive the expression of the evolution of the opinion offsets averaged over the noise and the sequences of interactions. The expression of $\overline{x_{ii}}(t+1)$ is the same as without gossip (equation \ref{eq:avavm}).

The expression of $\overline{x_{ji}}(t+1)$ includes an additional sum representing the average effect of agents gossiping with $j$ about $i$.
\begin{dmath}
 \label{eq:avavyg}
         \overline{x_{ji}}(t+1) = \overline{x_{ji}}(t) + \frac{2}{N_c} \left( \widehat{h_{ji}}(t)\left(\overline{x_{ii}}(t) - \overline{x_{ji}}(t)\right)  \\ + \overline{h'_{ji}}(t)\left(\overline{x_{ji}^2}(t) -  \overline{x_{ii}(t).x_{ji}(t)}\right)\right)  \\ + \frac{2k}{N_T} \sum_{p \notin \{i,j\}}   \widehat{h_{jp}}(t)\left(\overline{x_{pi}}(t) - \overline{x_{ji}}(t) \right),
\end{dmath}
where $N_T = N_a(N_a-1)(N_a-2)$.
The equations of $\overline{x_{ii}^2}(t+1)$, $\overline{x^2_{ij}}(t+1)$, $\overline{x_{ii}(t+1).x_{ji}(t+1)}$, $\overline{x_{ji}(t+1).x_{pi}(t+1)}$  for $(i,j, p) \in \{1,\dots,N_a\}^2$ are specified in the appendix (section \ref{sec:gosav}). Again, using the values of the terms at $t=0$, we can compute the values of $\overline{x_{ii}}(t)$ and $\overline{x_{ij}}(t)$ at any time step $t$ by induction.

The expressions of the positive bias on self-opinion and negative biases on the opinions about $i$ are the same as when there is gossip. Introducing the equilibrium opinion helps to understand how the biases are combined in the interactions and the impact of gossip.

\subsection{Interpretation of the equations and first order equilibrium opinion.}
\label{sec:equilibrium}

With or without gossip, at $t = 2$, the expressions of $\overline{x_{ii}}(2)$ and $\overline{x_{ji}}(2)$ are:
\begin{align} \label{eq:biast2ng}
    \overline{x_{ii}}(2) &= - \frac{4}{N_c^2} \left(\sum_{j \neq i} h'_{ij}(0)\right)\left(\sum_{j \neq i} h^2_{ij}(0)\right) \frac{\delta^2}{3},\\
    \overline{x_{ji}}(2) &=  \frac{4}{N_c^2} h'_{ij}(0)\left(1 - h_{ij}(0)\right)^2 \frac{\delta^2}{3}, \text{ for } j \neq i.
\end{align}

Like in the simplified case of only one varying opinion presented in section \ref{sec:biases}, $ \overline{x_{ii}}(2)$ is positive and  $\overline{x_{ji}}(2)$ is negative (because $h'_{ij}(0)$ is negative). Therefore, there is also a positive bias on the self-opinions and negative bias on the opinions about other, at step 2, whatever the number of interacting agents. Note that, because of the sums in the expression of $\overline{x_{ii}}(2)$, the positive bias is higher than the negative bias in absolute value, and this difference increases with the number of agents. Finally, both $\overline{x_{ii}}(2)$ and $\overline{x_{ji}}(2)$ are multiplied by $\frac{1}{N_c^2}$ indicating that the effect of the biases in one interaction decreases very strongly (in about $\frac{1}{N_a^4}$) when $N_a$ increases.

More generally, for $t >2$, let us consider the term of second order in equation \ref{eq:avavm} (same equation with or without gossip):
\begin{align}
   \frac{2}{N_c} \sum_{j \neq i}\overline{h'_{ij}}(t)\left(\overline{x_{ii}(t).x_{ji}(t)} - \overline{x_{ii}^2}(t) \right).
\end{align}
This term is positive as the derivative is assumed strictly negative. Therefore, the effect of this term is to increase the self-opinions and it can be seen as a positive bias on self-opinions. Similarly, the term of second order in equations \ref{eq:avavy} and \ref{eq:avavyg}:
\begin{align}
     \frac{2}{N_c}\overline{h'_{ij}}(t)\left(\overline{x_{ji}^2}(t) - \overline{x_{ii}(t).x_{ji}(t)}\right),
\end{align}

is negative and tends to decrease the opinion offset $x_{ji}$. It can be interpreted as a negative bias on opinions about others.

 Then, consider the terms of first order in equations \ref{eq:avavm} and \ref{eq:avavy} respectively:
\begin{align}
   &\frac{2}{N_c} \sum_{j \neq i}\widehat{h_{ij}}(t)\left(\overline{x_{ji}}(t) - \overline{x_{ii}}(t) \right),\\
   &\frac{2}{N_c} \widehat{h_{ji}}(t)\left(\overline{x_{ii}}(t) - \overline{x_{ji}}(t) \right), \mbox{ for } j \neq i.
\end{align}
 The effect of these terms is that opinions about $i$ attract each other, which keeps them close to each other. Therefore the positive and negative biases are combined into a common trend shared by all opinions about $i$.

 This common trend can be expressed by the first order equilibrium opinion offset $e_i(t)$ of agent $i$, or equilibrium opinion for short, which is defined as follows:
\begin{align}
    e_i(t) = \frac{1}{1+ S_i(t)}\left(\overline{x_{ii}}(t) +  \sum_{j\neq i}\frac{\widehat{h_{ij}}(t)}{\widehat{h_{ji}}(t)} \overline{x_{ji}}(t) \right),
\end{align}
with:
\begin{align}
    S_i(t) = \sum_{j\neq i}\frac{\widehat{h_{ij}}(t)}{ \widehat{h_{ji}}(t)}.
\end{align}
Indeed, when there is no gossip, applying equations \ref{eq:avavm} and \ref{eq:avavy} yields:
\begin{dmath} \label{eq:ei}
     e_i(t+1) = e_i(t) + \frac{2}{N_c(1+ S_i(t))}\sum_{i \neq j}\overline{h'_{ji}}(t)\left(\overline{x_{ii}(t).x_{ji}(t)} - \overline{x_{ii}^2}(t) + \frac{\widehat{h_{ij}}(t)}{\widehat{h_{ji}}(t)} \left(\overline{x_{ji}^2}(t) -  \overline{x_{ii}(t).x_{ji}(t)}\right)\right).
\end{dmath}

At any time step $t$, $e_i(t)$ is the value that would be reached by all the opinions about $i$ if the $h_{ij}(t)$ were frozen. More precisely, imagining that from a given time $t_0$, for all $t > t_0$ and for all $(i,j) \in \{1, \dots, N_a\}$, $h_{ij}(t) = h_{ij}(t_0)$, then $\overline{x_{ji}}(t)$ for all $j$ would converge to $e_i(t_0)$ and remain at this value. Therefore, the term of second order in equation \ref{eq:ei} determines the second order effect applied to an opinion which is at the equilibrium of the first order effects. In the long run, this trend is common to all opinions about $i$, as the opinions about $i$ reach their equilibrium distances from each other (see trajectory examples on Figure \ref{fig:xjitN10e12}).

The trend is thus expressed as a weighted sum of the positive bias on the self-opinion and the negative biases on the opinions about $i$. The negative biases are multiplied by the factor $\frac{\widehat{h_{ij}}(t)}{\widehat{h_{ji}}(t)}$, which is high when $\overline{a_{ii}} < \overline{a_{ij}}$ and $\overline{a_{jj}} > \overline{a_{ji}}$. Therefore, when the agents are in a consensual hierarchy, these factors are higher for agents $i$ of low status. Hence, the opinions about the agents of low status grow less (or even can decrease) than the opinions about the agents of high status.

When gossip is activated, the equation of $e_i(t+1)$ remains the same except that a term of first order coming from gossip is added.
However, simulations show that the effect of this term is negligible. Therefore, like in the case without gossip,  $e_i(t)$ provides the common trend of the evolution of the opinions about $i$. 

However, the additional term accounting for gossip modifies the negative bias on the opinion about others.
Indeed, we have:
\begin{dmath}
\label{eq:y2gossip}
 \overline{x^2_{ji}}(t+1) = \overline{x^2_{ji}}(t) + \frac{2}{N_c} \left(\overline{G^2_{ji}}(t) - \overline{x_{ji}^2}(t) + \overline{h_{ji}}^2(t) \frac{\delta^2}{3}\right) + \frac{2k}{N_T} \sum_{p \notin \{j,i\}} \left( \overline{J^2_{jpi}}(t) - \overline{x_{ji}^2}(t) + \overline{h_{jp}}^2(t) \frac{\delta^2}{3}\right),
\end{dmath}
with:
\begin{align}
  \overline{G_{ji}}(t)  &= \overline{x_{ji}(t)} + \widehat{h_{ji}}(t)\left(\overline{x_{ii}}(t)-\overline{x_{ji}}(t) \right), \\
   \overline{J_{jpi}}(t) &= \overline{x_{ji}}(t) + \widehat{h_{ji}}(t)\left( \overline{x_{pi}}(t) -  \overline{x_{ji}}(t)\right).
\end{align}

This additional sum increases $\overline{x_{ji}^2}(t+1)$, which increases the negative bias on $x_{ji}$ in the following time steps. 
In particular, at time $t = 2$, $\overline{x_{ii}}(2)$ has the same expression as in equation \ref{eq:biast2ng}, and we have, for $i \neq j$:
\begin{align} \label{eq:biast2g}
    \overline{x_{ji}}(2) &=  \frac{4}{N_c^2} h'_{ij}(0)(1 - h_{ij}(0))^2 \frac{\delta^2}{3} + \frac{4}{N_c N_T} h'_{ij}(0)\left( \sum_{p \notin \{i,j\}} h_{jp}^2(0)\right)  \frac{\delta^2}{3},
\end{align}

When the agents are in a consensual hierarchy, the additional negative bias is stronger for $i$ of low status, because $\overline{h_{jp}}(t)$ is higher for $j$ of low status and $\overline{h'_{ij}}(t)$ is higher when the statuses of $i$ and $j$ are close. This explains a stronger negative effect of gossip on agents of low status.

\section{Numerical experiments}
In a first set of experiments, we check the accuracy of the moment approximation. In the second set of experiments, using the moment approximation, we investigate the effect of inequalities on the evolution of the opinions. 

\subsection{Accuracy of the moment approximation}

\subsubsection{Examples when only one opinion varies between two interacting agents.}
\begin{figure}
	\centering
	\begin{tabular}{m{1em}cm{1em}c}
	  \rotatebox{90}{$\overline{x_{11}}(t)$} &	\makecell{ \includegraphics[width= 6.0 cm]{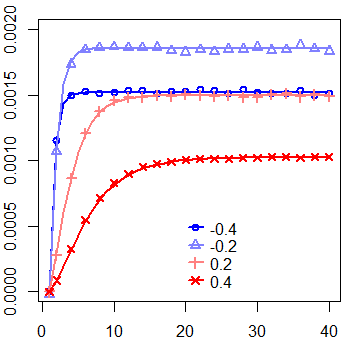} }  & \rotatebox{90}{$\overline{x_{21}}(t)$}	& \makecell{\includegraphics[width= 6.0cm]{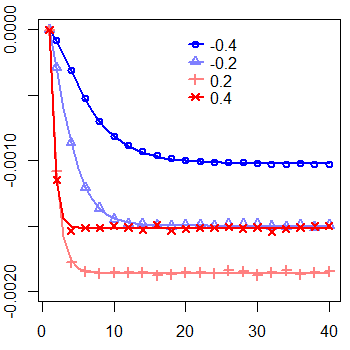}} \\
	  & $t$ & & $t$
	\end{tabular}
\caption{Biases when only one opinion is varying. Left panel: $\overline{x_{11}}(t)$, right panel: $ \overline{x_{21}}(t)$, for $a_{12}(t) = b = 0$. The different colours represent values of $a_{11}(0)$ (left panel) or $a_{21}(0)$ (right panel) -0.4, -0.2, 0.2, and 0.4, as specified in the legend. $t$ is the number of encounters. Each point is the average of the biases over 10 million simulations and the lines are computed with equation \ref{eq:posBiast} for the positive bias and its equivalent for the negative bias. Influence parameter $\sigma = 0.3$. Noise parameter $\delta = 0.1$.}
\label{fig:biasest2}
\end{figure}

We first check the accuracy of the approximation in the simplified setting of section \ref{sec:biases} where only $x_{11}$ or $x_{21}$ is varying.
Figure \ref{fig:biasest2} shows the value of $\overline{x_{11}}(t)$ and $ \overline{x_{21}}(t)$ from the theoretical formulas (equation \ref{eq:posBiast} and its transformation for the negative bias) and from the average of 10 million repetitions of the simulation during 40 encounters. The value of $a_{12}(t) = b = 0$ is fixed and the curves corresponding to four different values of $a_{11}(0) = a_{21}(0) = a$ are shown in different colours in the graphs (see legend). The approximation appears very accurate.

\subsubsection{Examples of trajectories of opinions about an agent for 10 interacting agents, without gossip}

\begin{figure}
	\centering
	\begin{tabular}{m{1em}cc}
	&$a_{ii}(0) = 0.6$ (status = 10) & 	$a_{ii}(0) = 0.47$ (status = 9)\\
	  \rotatebox{90}{ $\overline{x_{ji}}(t)$} &  	\makecell{\includegraphics[width= 6 cm]{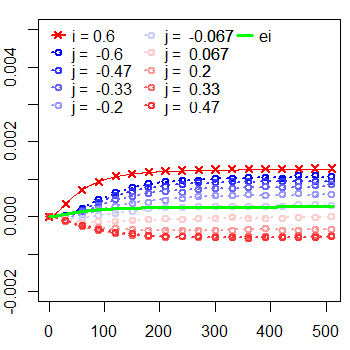}}  & \makecell{\includegraphics[width= 6 cm]{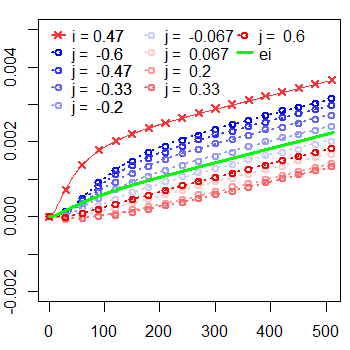}} \\
	 & $t$ & $t$\\[15pt]
	 &$a_{ii}(0) = -0.47$ (status = 2) & 	$a_{ii}(0) = -0.6$ (status = 1)\\
  \rotatebox{90}{  $\overline{x_{ji}}(t)$} &  	\makecell{\includegraphics[width= 6 cm]{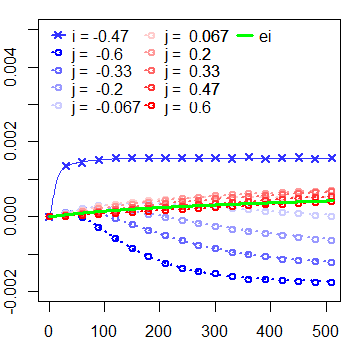}}  & \makecell{\includegraphics[width= 6 cm]{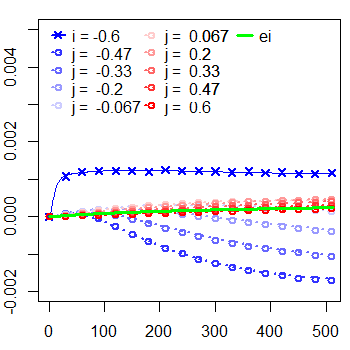}}\\
 & $t$ & $t$
	\end{tabular}
	\caption{Examples of evolution of average opinion offsets $ \overline{x_{ji}}(t)$ for 10 agents without gossip, with $a_{ii}(0) \in [-0.6, 0.6]$.  The lines are obtained with the moment approximation and the points by averaging the results of 10 million simulations. The legend indicates the colour corresponding to the rank of the initial self-opinion $a_{ii}(0)$. The equilibrium opinion $e_i(t)$ is represented in green. Noise parameter $\delta = 0.1$. Influence function parameter $\sigma = 0.3$. Number of gossip $k=0$.}
	\label{fig:xjitN10e12}
\end{figure}

Figure \ref{fig:xjitN10e12} shows examples of the trajectories of $\overline{x_{ji}}(t)$ for $j \in \{1, \dots, N_a\}$, for a given agent $i$. The title above each panel specifies the value of $a_{ii}(0)$. The lines (solid for the self-opinions $\overline{x_{ii}}(t)$, dashed for the opinions $\overline{x_{ji}}(t)$ with $j \neq i$) are obtained by the moment approximation while the points are the average values of 10 million simulations. The accuracy seems quite satisfactory.

The trajectory of the equilibrium opinion, computed with the moment approximation (see section \ref{sec:equilibrium}), is represented in green. In the top panels, $i$ is of high status (high values of $a_{ii}(0)$) and from $t = 300$, all the trajectories grow with a very similar slope. This is not the case for the bottom panels where $i$ is of low status (low values of $a_{ii}(0)$). Indeed, the trajectories are increasing for $j$ of high status (red shades) while they are decreasing for $j$ of low status (blue shades). The trajectory of the equilibrium opinion (in green) tends to be closer to the trajectories of $\overline{x_{ji}}(t)$ for $j$ of high status (red shades). For larger values of $t$ however (not visible on the graph), all the trajectories become progressively almost parallel. Moreover, in all cases in our simulations, for all $j \neq i$ and for all $t$, we have:  $\overline{x_{ii}}(t) > \overline{x_{ji}}(t)$.

\subsubsection{RRMSE of the moment approximation for different number of agents.}
In order to evaluate more quantitatively the accuracy of the moment approximation, we compute the root of the relative mean squared error (RRMSE)\footnote{The RRMSE is the root mean squared error divided by the mean of absolute value of the target values.} between the moment approximation and the average results over 10 million simulations. The initial opinions are all such that for all $(i,j)$, $x_{ji}(0) = x_{ii}(0)$, and $x_{ii}(0)$ are regularly distributed on the interval [-0.3, 0.3]. For any $t$ and any couple $(i,j)$, let $\overline{\overline{x_{ji}}}(t)$ be the average of $x_{ji}(t)$ over 10 million simulations. Keeping the notation $\overline{x_{ji}}(t)$ for the moment approximation,  $\mathcal{E}\left(\overline{x_{ji}}(1,\dots,T)\right)$, the RRMSE of $\overline{x_{ji}}(t)$ for $t \in [1, T]$ is:
\begin{align} \label{eq:RRMSE}
    \mathcal{E}\left(\overline{x_{ji}}(1,\dots,T)\right) = \frac{\sqrt{T\left(\sum_{t = 1}^{T}\left( \overline{x_{ji}}(t) - \overline{\overline{x_{ji}}}(t)\right)^2\right)}}{\sum_{t = 1}^{T} |\overline{\overline{x_{ji}}}(t)|}.
\end{align}
 We define the RRMSE for the second moment variables similarly.
 
\begin{figure}
	\centering
	\begin{tabular}{m{1em}cc}
	&No Gossip ($k = 0$) & Gossip ($k = 1$)\\
	  \rotatebox{90}{ First Moment} &  	\makecell{\includegraphics[width= 6 cm]{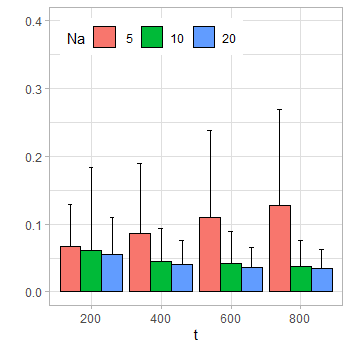}}  & \makecell{\includegraphics[width= 6 cm]{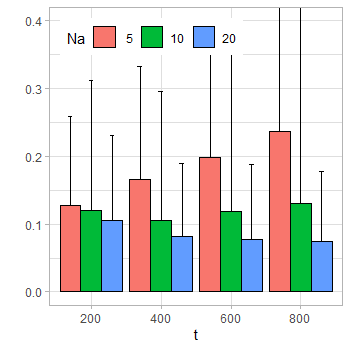}} \\[-5pt]
	  \rotatebox{90}{ Second Moment} &  	\makecell{\includegraphics[width= 6 cm]{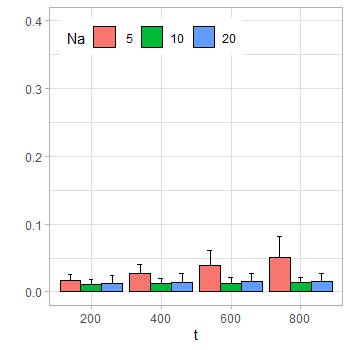}}  & \makecell{\includegraphics[width= 6 cm]{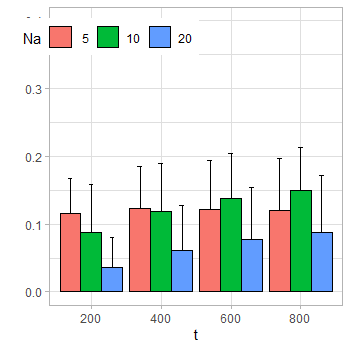}}\\[-5pt]
	\end{tabular}
	\caption{Average RRMSE of the approximation (defined by equation \ref{eq:RRMSE}). Top panels: average for all first moment variables ($\overline{x_{ii}}(t)$ and $\overline{x_{ji}}(t)$). Bottom panels average for all second moment variables ($\overline{x^2_{ii}}(t)$, $\overline{x^2_{ji}}(t)$ and $\overline{x_{ii}(t)x_{ji}(t)}$) for a number of agents $N_a \in \{5, 10, 20\}$. The RRMSE is computed on the interval $[1,t]$, $t$ being defined on the horizontal axis. The error bars show the standard deviations in the considered set of variables ($N_a^2$ variables for the first moment, $N_a^3$ variables for the second moment). Noise parameter $\delta = 0.1$. Influence function parameter $\sigma = 0.3$.}
	\label{fig:RRMSE}
\end{figure}

 Figure \ref{fig:RRMSE} shows the average RRMSE computed for the first moment and the second moment variables, for a number of agents $N_a \in \{5, 10, 20\}$. The RRMSE is on average lower than 10 \% for the dynamics without gossip and $N_a > 5$. With gossip, the RRMSE is higher, but it is lower than 15 \% when  $N_a > 5$. It can be expected that the approximation, neglecting the terms of degree higher than 2, is more accurate while the values of $x_{ji}$ and $x_{ji}^2$ remain small. When the number of agents increases, we have seen that $\overline{x_{ji}}(t)$ is multiplied by a factor of order $\frac{1}{Na^4}$, therefore it can be expected that the error gets smaller for the same values of $t$, when $N_a$ increases. Similarly, gossip increases $x_{ji}^2(t)$, which is expected to decrease the approximation accuracy at $t$.

\subsection{Effect of initial inequalities on the evolution of the opinions.}
In the following experiments, we illustrate the effect of initial inequalities on the evolution of opinions in the short term (a few hundreds of interactions) on the case of 10 agents. Indeed, this number of agents is low enough for readable exhaustive representations and high enough for a reasonably accurate moment approximation (RRMSE < 15\% on average). The initial opinions are the same in each column of the opinion matrix, and the initial self-opinions (that equal the initial reputations) are regularly distributed in an interval $[-w, w]$. Increasing $w$ corresponds to increasing inequalities. The status of the agent of the highest initial self-opinion is 10 (highest status) and the status of the agent of initial lowest self-opinion is 1 (lowest status).

\subsubsection{Comparing trajectories of equilibrium opinion for two different inequality widths}

\begin{figure}
	\centering
	\begin{tabular}{m{1em}cc}
	&No Gossip ($k = 0$) & Gossip ($k = 1$)\\
	  \rotatebox{90}{ $e_i(t)$} &  	\makecell{\includegraphics[width= 6 cm]{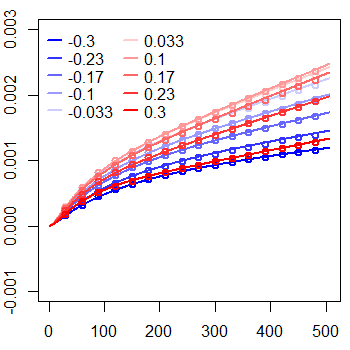}}  & \makecell{\includegraphics[width= 6 cm]{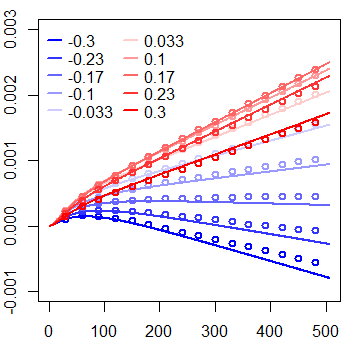}} \\
	  \rotatebox{90}{ $e_i(t)$} &  	\makecell{\includegraphics[width= 6 cm]{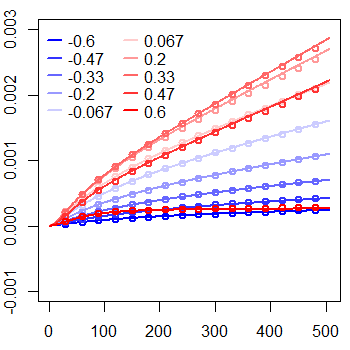}}  & \makecell{\includegraphics[width= 6 cm]{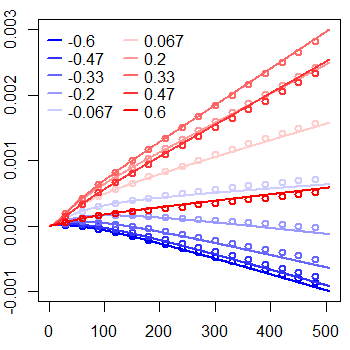}}
	\end{tabular}
	$t$ (pair encounters)
	\caption{Evolution of equilibrium opinions ($ \overline{r_i}(t)$) for 10 agents with $a_{ii}(0) \in [-0.3, 0.3]$ (top panels) or $a_{ii}(0) \in [-0.6, 0.6]$ (bottom panels). The lines are obtained with the moment approximation and the points by averaging the results of 10 million simulations. The legend provides the colour corresponding to the initial self-opinion $a_{ii}(0)$. Noise parameter $\delta = 0.1$. Influence function parameter $\sigma = 0.3$.}
	\label{fig:reputsN10}
\end{figure}

Figure \ref{fig:reputsN10} represents the average evolution of equilibrium opinions $e_i(t)$ for $i \in \{1,\dots, 10\}$ during 500 pair encounters when the starting self-opinions are uniformly distributed in [-0.3, 0.3] or in [-0.6, 0.6] and with or without gossip. The main features shown by this figure are the following:
\begin{itemize}
    \item In the top left panel, with small inequalities and without gossip, all equilibrium opinions are increasing and remain close to each other. 
    \item In the top right panel, with small inequalities and with gossip, the two lowest equilibrium opinions are slightly decreasing. The trajectories of highest status agents are similar with or without gossip.
    \item In the left bottom panel, with large inequalities and without gossip, the trajectories for agents of high status (in shades of red) increase like when inequalities are low, except for the agent of top status which increases more slowly. However, for the agents of lower status (in shades of blue), the trajectories increase significantly less than when inequalities are low;
    \item In the bottom right panel, with large inequalities and with gossip, the trajectories for agents of high status are similar to the ones without gossip. However, for the four agents of lowest status, the trajectories are significantly different; they are decreasing (in blue).
\end{itemize}

\subsubsection{Slope of equilibrium opinion trajectory at after 800 encounters when inequalities vary}
\begin{figure}
	\centering
	\begin{tabular}{m{1em}cc}
	&No Gossip ($k = 0$) & Gossip ($k = 1$)\\
	\rotatebox{90}{ $e_i(800)-e_i(799)$} &  	\makecell{\includegraphics[width= 6 cm]{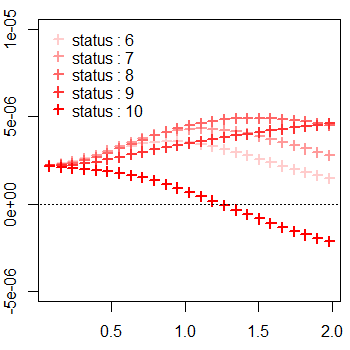}}  & \makecell{\includegraphics[width= 6 cm]{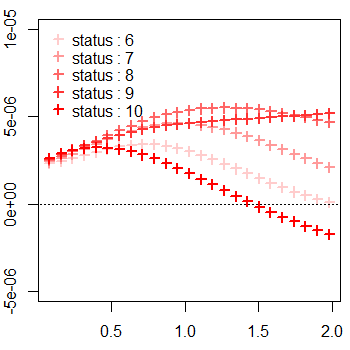}}\\
	\rotatebox{90}{ $e_i(800)-e_i(799)$} &  	\makecell{\includegraphics[width= 6 cm]{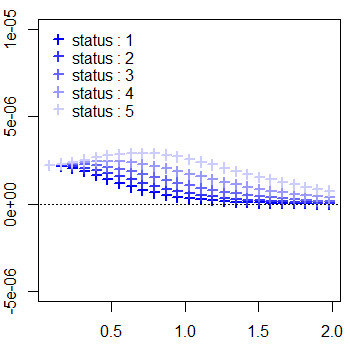}}  & 	\makecell{\includegraphics[width= 6 cm]{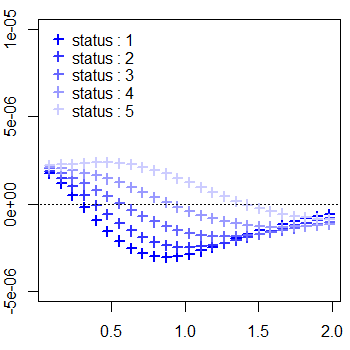}}
	\end{tabular}
	Width of initial self-opinions interval (inequalities)
	\caption{Slope of equilibrium opinion trajectory at $t = 800$ ( $e_i(800)-e_i(799)$) computed by the moment approximation, for 25 different initial ranges of opinions (horizontal axis). The colour of the points codes for the status of the agent; the top panels represent the high statuses and the bottom panels the low statuses. On the left panels, there is no gossip, on the right panels there is ($k = 1$). Noise parameter $\delta = 0.1$. Influence function parameter $\sigma = 0.3$.}
	\label{fig:influSo}
\end{figure}

In each panel of Figure \ref{fig:influSo}, the x-axis represents the width of the initial opinion intervals varying from [-0.06, 0.06] to [-0.9, 0.9], the curves represent $e_i(800)-e_i(799)$, the slope of the trajectory of the equilibrium opinion at $t = 800$, computed with the moment approximation. The colour of the curve codes for the status of agent $i$. In general, this slope is close to the slopes of the opinions about agent $i$.

This figure shows that:
\begin{itemize}
    \item For agents $i$ of high status (in red, top panels):
    \begin{itemize}
        \item The left and right panels are similar, except for the agent of status 6 for which the equilibrium opinion shows a significantly lower slope with gossip, when inequalities increase;
        \item The slope for the agent of the highest status decreases when the inequalities are above a threshold, while the slopes  for all the other agents of high status remain positive, both with and without gossip;
        \item The agents of the highest statuses (7 and above) have a slightly higher slope when there is gossip:
    \end{itemize}
    
    \item For agents $i$ of low status (in blue, bottom panels): 
    \begin{itemize}
        \item When there is no gossip (left panel), all the slopes remain positive, but they all tend to 0 when the inequalities increase;
        \item When there is gossip (right panel), the slopes become negative as the inequalities increase. The slope for the agent of the lowest status becomes negative first, then the slope for the agent for second lowest status becomes negative and so on until the slopes of all the agents of low status become negative. 
    \end{itemize}

\end{itemize}

\section{Discussion}
\subsection{Relevance of the moment approximation}
The moment approximation appears reliable while the number of agents is higher or equal to 10 and the number of encounters remains below 1000. Moreover, it provides explanations of the model behaviour. 
\begin{itemize}
\item After a while, the opinions about an agent tend to evolve in parallel, and their evolution is driven by a second order effect, which is a weighted sum of a positive bias on self-opinions and negative biases on the opinions about others; 
\item In absolute value, the positive bias at $t = 2$ $\overline{x_{ii}}(2)$ is higher than the negative biases at $t = 2$  $\overline{x_{ji}}(2)$ as soon as the number of agents is higher than 2 and the difference increases when the number of agents increases.  Indeed, all  encounters of agent $i$ contribute to the average positive bias on $i$'s self-opinion while the negative bias on the opinion of $i$ about a particular agent $j$ is generated only when $i$ interacts with $j$. Moreover, the values $\overline{x_{ii}}(2)$ and $\overline{x_{ji}}(2)$ are generated by the noise at each iteration $t$ and added to the biases at $t$. This suggests that, without gossip, the positive bias tends to dominate;
\item The weights on the negative biases are higher for the agents of low status and these weights increase when the inequalities increase. This explains why the opinions about the low status agents tend to stagnate or decrease (when there is gossip);
\item When there is gossip, a term is added to  $\overline{x_{ji}}(t)$, which increases $\overline{x_{ji}^2}(t)$ and the negative bias on the opinion about others. Moreover, this added term is stronger when $i$ and $j$ are of low status and quite small for agents $i$ of high status, especially when the initial inequalities are high. This explains the observed higher negative effect of gossip on the agents of low status.
\end{itemize}
However, we could not explain mathematically that the evolution of $\overline{x_{ii}}(t)$  and all  $\overline{x_{ji}}(t)$ become progressively parallel, and for all $t$,  $\overline{x_{ii}}(t) >\overline{x_{ji}}(t)$  for all $j \neq i$, which expresses a more standard definition of positive bias. 

In addition to the explanations that mathematical expressions can bring, the moment approximation provides a means to explore the average behaviour of the agent based model, without running millions of simulations. Such an exploration for 10 agents, when varying the width of the interval of initial self-opinions, reveals the following features (see figure \ref{fig:influSo}):
\begin{itemize}
\item The opinions about the agents of high status (except the top status) tend to grow in roughly the same way, with or without gossip;
\item  Without gossip, the opinions about agents of low status tend to grow but this growth progressively decreases and even becomes close to zero when the initial inequalities increase. With gossip, these opinions grow only when the inequalities are very low and then the opinions about low status agents start decreasing as the inequalities increase.
\end{itemize}
The same explorations conducted with 20 and 40 agents yield similar results.

These observations provide some explanations to the patterns recalled in section \ref{sec:patterns}. Indeed, initially, in these patterns, all the opinions are the same, therefore, both with and without gossip, all the average opinions tend to grow together in a first period of a few thousand steps. However, because of the noise, more or less dispersion of the opinions takes place, introducing inequalities between agents:
\begin{itemize}
\item Without gossip, since all opinions tend to grow at a similar pace, the opinion inequalities remain moderate for a while and all opinions grow on average. When the inequalities reach a threshold though, the opinions about agents of low status grow more and more slowly or stagnate, while the opinions about agents of high status fluctuate when reaching the opinion limit at +1. Overall, the distribution of opinions is therefore significantly positive on average;
\item When there is gossip, because the opinions about agents of low status grow much more slowly than the opinions about agents of high status, the inequalities of opinions increase more rapidly and easily reach a level in which the opinions about the lowest status agent starts decreasing, which further increases the opinion inequalities, and the opinions about other agents of low status start decreasing, which further increases the opinion inequalities. Ultimately, when the inequalities are maximum, the opinions about a majority of agents tend to decrease. This explains why the overall distribution of opinions becomes negative on average.
\end{itemize}

We checked the validity of these explanations by introducing a process that limits the inequalities of the opinions by regularly driving them slightly towards their average. More precisely, every $N_a$ interactions, all opinions are modified, using parameter $\lambda$, as follows:

\begin{align}  \label{eq:grouping}
    a_{ij}(t+1) &= (1-\lambda)a_{ij}(t) + \lambda. \overline{a}, \forall i,j \in N_a,
\end{align}
with $\overline{a}$ being the average opinion. With this modified dynamics, the opinions grow and stabilise to a high average positive value, even when parameter $\lambda$ is small (0.0001) and when there is gossip.
\subsection{Connections with the literature in social-psychology}

We now discuss how the model behaviour relates to some researches in social-psychology. We first consider the biases and then we discuss the effect of inequalities on the evolution of opinions. 

Social-psychology robustly established that people tend to overestimate themselves (for instance overoptimism or overconfidence in judgement and predictions or the ability to complete a task or about forecasting events in general, see \cite{Dunning2004} for a review). This tendency is often called positivity bias. A widely accepted explanation relates the positivity bias to the well established tendency of most people to self-enhancement or self-protection. People tend to seek out and accept positive feed-backs and to avoid or reject negative ones \cite{Sedikides1997}. Indeed, when we receive a negative feedback, we often tend to decrease our evaluation of its source and thus we decrease its importance (e.g. \cite{Campbell1999}). As a result, on average, negative feed-backs tend to have a lower impact than positive ones on self-evaluation, which leads to self-overestimation \cite{Moreland1984}. This process presents strong similarities with the positive bias observed in the Leviathan model, when vanity is active, because vanity decreases the evaluation of the source of negative feed-backs (see details in \cite{Deffuant2013}). 

However, the positive bias studied in this paper is generated by the model without vanity and suggests the existence of another mechanism. Indeed, this positive bias cannot be attributed to any self-enhancement. It is a statistical effect of the noise combined with the decreasing influence function. As far as we know, this specific bias has not been observed by social-psychologists.

Considering the negative bias now, the literature reports some negative tendencies in judging others. People show a negative bias on the opinion about others in some specific contexts, for instance when requested to express their opinion in front of an audience of higher status (see for instance \cite{Amabile1981}). Also, when judging moral qualities of others, we tend to put a higher weight on the negative features than on the positive ones (while this is the opposite when judging the abilities) \cite{Martijn1992,Skowornski1992}. However, the contexts of these observations are difficult to relate to our model.

Therefore, it seems that, if they do exist in human interactions, the biases observed in our model have been overlooked by social-psychologists. This would not be surprising, since these biases are small (of second order). Moreover, though our simulations suggest that their long term effect is potentially huge, it is impossible to relate these effects to their causes without the type of analysis that we carried out. 

Now, we consider possible connections between the literature in social-psychology and the effect of inequalities observed in the model. 

A study involving a cohort of  3058 adolescents in Denmark, followed from ages 15 to 21, shows that the self-esteem of the adolescents from the richest tertile grows significantly more than the self-esteem of the poorest tertile \cite{Hansen2014}. This result seems in line with our model patterns. However, the mechanisms involved are probably quite different. Indeed, the impact of inequalities on self-esteem is generally related to personal or group self-deprivation or feeling of injustice \cite{Walker1999,Crocker1999}. These feelings are absent from the model.  Again, the model shows a statistical phenomenon of second order, while the reaction to self-deprivation, that could probably be also modeled using the vanity process of the Leviathan model, is certainly of first order. Hence, again, it seems very likely that the research in social psychology missed the phenomenon suggested by our model.

Since we cannot rely on existing literature to get adequate experimental data challenging the model, a solution is to get this data by running specifically designed experiments. The following directions can be envisaged: 

\begin{itemize}
\item The model suggests the existence of a positive bias on self-opinions and a negative bias on opinions about others, without self-enhancement or self-protection (i.e. with symmetric reactions of same intensity to negative and positive feed-backs of same intensity), as soon as the influence function is decreasing when the self-opinion is increasing. The main feature to check is thus this property of the influence function, because its presence mathematically implies the biases. It seems possible to design an experiment achieving this;
\item The model suggests that the interactions in a group with wide perceived inequalities tend to widen these perceived inequalities by decreasing the opinions about the agents of low status (especially when there is gossip) and by increasing the opinions about the agents high status. However, in a group with small perceived inequalities, the interactions tend to increase the opinions about all the agents, even if there is gossip. Therefore, in this perspective, introducing mechanisms that limit the perceived inequalities (like the mechanism described by equation \ref{eq:grouping}) in a group should be beneficial to the opinions about all its members. Experiments checking these predictions could consider a participant interacting with a set of experimenters and controlling the respective statuses of the participant and of the experimenters.
\end{itemize}

Overall, this work identifies second order effects of interactions that seem impossible to observe without suspecting their existence. However, it seems possible to design specifically targeted experiments that would detect them.

\section{Acknowledgement}

This work has been partly supported by the Agence Nationale de la Recherche through the ToRealSim project within the ORA program and the FuturICT2.0 projet within the Flagera program. We are grateful to Sylvie Huet for her comments on a preliminary version of the paper and to the whole ToRealSim project team for helpful discussions.

\section{Additional information}
The code of the model is available at: \url{https://www.comses.net/codebases/12d44111-5823-4773-ad59-754ebacb33a1/releases/1.0.0/}.

\bibliographystyle{splncs04}
\bibliography{references}

\section{Appendix}

\subsection{Equations of second moments for a given sequence of interactions, without gossip}
\label{sec:nogos}

Let $\widehat{h_{ij}}(s_t) = \overline{h_{ij}}(s_t) - \overline{h'_{ij}}(s_t)\overline{z_{ij}}(s_t)$.
For $(i, j) = (i_{t+1},j_{t+1})$, or  $(i, j) = (j_{t+1},i_{t+1})$:
\begin{dmath}
 \label{eq:avmapp}
      \overline{x_{ii}}(s_{t+1}) =  \overline{x_{ii}}(s_t) + \widehat{h_{ij}}(s_t)\left(\overline{x_{ji}}(s_t)-x_{ii}(s_t) \right) \\-  \overline{h'_{ij}}(s_t)\left(\overline{x^2_{ii}}(s_t) - \overline{x_{ii}(t)x_{ji}(s_t)} \right),
\end{dmath}

and:
 \begin{dmath}
\label{eq:avyapp}
      \overline{x_{ji}}(s_{t+1}) = \overline{x_{ji}}(s_t) + \widehat{h_{ji}}(s_t)\left( \overline{x_{ii}}(s_t)-\overline{x_{ji}}(s_t) \right) \\+  \overline{h'_{ji}}(s_t)\left(\overline{x^2_{ji}}(s_t) - \overline{x_{ii}(t)x_{ji}(s_t)}\right).
 \end{dmath}

Let:
\begin{align}
    \overline{F_{ij}}(s_t) = \overline{x_{ii}}(s_t) + \widehat{h_{ij}}(s_t)\left(\overline{x_{ji}}(s_t)-\overline{x_{ii}}(s_t) \right).\\
    \overline{G_{ji}}(s_t)  = \overline{x_{ji}(s_t)} + \widehat{h_{ji}}(s_t)\left(\overline{x_{ii}}(s_t)-\overline{x_{ji}}(s_t) \right).
\end{align} 

 Neglecting the terms of order higher than 2, we get:
\begin{dmath}
      \overline{x^2_{ii}}(s_{t+1}) = \overline{F_{ij}^2}(s_t)  + \overline{h_{ij}}^2(s_t) \frac{\delta^2}{3},
\end{dmath} 

with:

\begin{dmath}
   \overline{F_{ij}^2}(s_t) = (1- \widehat{h_{ij}}(s_t))^2 \overline{x_{ii}^2}(s_t) + \widehat{h_{ij}}^2(s_t)\overline{x_{ji}^2}(s_t) \\
   + 2 (1- \widehat{h_{ij}}(s_t))\widehat{h_{ij}}(s_t) \overline{x_{ii}(s_t)x_{ji}(s_t)},
\end{dmath}

and:
\begin{dmath}
       \overline{x^2_{ji}}(s_{t+1}) = \overline{G_{ji}^2}(s_t) + \overline{h_{ji}}^2(s_t) \frac{\delta^2}{3},
\end{dmath} 

with:
\begin{dmath}
   \overline{G_{ji}^2}(s_t) = (1- \widehat{h_{ji}}(s_t))^2 \overline{x_{ji}^2}(s_t) + \widehat{h_{ji}}^2(s_t)\overline{x_{ii}^2}(s_t) \\
   + 2 (1- \widehat{h_{ji}}(s_t))\widehat{h_{ji}}(s_t) \overline{x_{ii}(s_t)x_{ji}(s_t)}.
\end{dmath}

Similarly, we get:
 \begin{dmath}
     \overline{x_{ii}(s_{t+1}).x_{ji}(s_{t+1})} = \overline{F_{ij}(s_t)G_{ji}(s_t)}.
\end{dmath}

For $p \neq i$ and $p \neq j$:

\begin{dmath}
     \overline{x_{ii}(s_{t+1}).x_{pi}(s_{t+1})} = \overline{F_{ij}(s_t)x_{pi}(s_t)}.
\end{dmath}

\begin{dmath}
     \overline{x_{ji}(s_{t+1}).x_{pi}(s_{t+1})} = \overline{G_{ji}(s_t)x_{pi}(s_t)}.
\end{dmath}

\subsection{Equations of second moments for all sequences of interactions, without gossip}
\label{sec:nogosav}

For any $i \in \{1,\dots, N_a\}$:

\begin{dmath}
 \label{eq:avavmsq3}
      \overline{x^2_{ii}}(t+1) =  \frac{N_a-2}{N_a}\overline{x^2_{ii}}(t) + \frac{2}{N_c} \sum_{j \neq i} \left(\overline{F_{ij}^2}(t) + \overline{h_{ij}}(t)^2 \frac{\delta^2}{3}\right).
\end{dmath}

For $j \neq i$:
 \begin{dmath}
 \label{eq:avavy3}
      \overline{x^2_{ji}}(t+1) = \frac{N_c-2}{N_c} \overline{x^2_{ji}}(t) +\frac{2}{N_c}\left(\overline{G_{ji}^2}(t) + \overline{h_{ji}}^2 \frac{\delta^2}{3}\right).
\end{dmath}

Moreover:
 \begin{dmath}
     \overline{x_{ii}(t+1).x_{ji}(t+1)} = \frac{N_a-2}{N_a}\overline{x_{ii}(t).x_{ji}(t)} +\frac{2}{N_c} \left( \overline{F_{ij}(t)G_{ji}(t)} + \sum_{p \notin \{i,j\}} \overline{F_{ip}(t)x_{ji}(t)} \right).
\end{dmath}

For $(i, j, p) \in \{1,\dots, N_a\}^3$, $i \neq j$, $j \neq 
p$, $i \neq p$:

\begin{dmath}
     \overline{x_{ji}(t+1).x_{pi}(t+1)} =  \frac{N_c - 4}{N_c} \overline{x_{ji}(t).x_{pi}(t)} +\frac{2}{N_c} \left( \overline{x_{ji}(t)G_{pi}(t)} +  \overline{x_{pi}(t)G_{ji}(t)}\right),
\end{dmath}

with, for instance:

\begin{dmath}
     \overline{F^2_{ij}}(t) = (1- \widehat{h_{ij}}(t))^2 \overline{x_{ii}^2}(t) + \widehat{h_{ij}}^2(t)\overline{x_{ji}^2}(t) + 2 (1- \widehat{h_{ij}}(t))\widehat{h_{ij}}(t) \overline{x_{ii}(t)x_{ji}(t)}.
\end{dmath}

 Starting from the values at $t = 0$, applying these equations, we can compute $\overline{x_{ii}}(t)$ and $\overline{x_{ij}}(t)$.

\subsection{Equations of second moments for a given sequence,  with gossip}
\label{sec:gos}
For $(i,j) = (i_{t+1}, j_{t+1})$, all the products are the same as the ones in the case without gossip.

Moreover, for $g \in \{g_{1_t},\dots,g_{k_t}\}$, let: 
\begin{dmath}
       \overline{J_{ijg}}(s_{t}) = \overline{x_{ig}}(s_{t}) + \widehat{h_{ij}}(s_t)\left( \overline{x_{jg}}(s_{t}) -  \overline{x_{ig}}(s_{t})\right).
\end{dmath}
We have:
\begin{dmath}
   \overline{x^2_{ig}}(s_{t+1}) = \overline{J_{ijg}^2}(s_t) +  \overline{h_{ij}}^2(t) \frac{\delta^2}{3}.
\end{dmath}
Moreover:
\begin{dmath}
   \overline{x_{ig}(s_{t+1})x_{jg}(s_{t+1})} = \overline{J_{ijg}(s_t)J_{jig}(s_t)}.
\end{dmath}

\subsection{Equations of second moments for all sequences of interactions, with gossip}
\label{sec:gosav}
For $(i,j) \in \{1,\dots,N_a\}^2$, like without gossip we have :
\begin{dmath}
      \overline{x^2_{ii}}(t+1) = \frac{N_a-2}{N_a} \overline{x^2_{ii}}(t) + \frac{2}{N_c} \sum_{j \neq i} \left(\overline{F_{ij}^2}(t) + \overline{h_{ij}}^2(t) \frac{\delta^2}{3}\right).
\end{dmath}
However, $\overline{x^2_{ji}}(t+1)$ is different with gossip:
\begin{dmath}
     \overline{x^2_{ji}}(t+1) = P_1.\overline{x^2_{ji}}(t) + \frac{2}{N_c} \left(\overline{G_{ji}^2}(t) + \overline{h_{ji}}^2(t) \frac{\delta^2}{3}\right) + \frac{2k}{N_T} \sum_{p \notin \{j,i\}} \left( \overline{J_{jpi}^2}(t) + \overline{h_{jp}}^2(t) \frac{\delta^2}{3}\right),
\end{dmath}

with:
\begin{align}
    N_T &= N_a(N_a-1)(N_a -2),\\
    P_1 &= 1 - \frac{2}{N_c} - \frac{2k}{N_c}.
\end{align}

The average products $\overline{x_{ii}(t+1).x_{ji}(t+1)}$  and $\overline{x_{ji}(t+1).x_{pi}(t+1)}$ are also different with gossip:

 \begin{dmath}
     \overline{x_{ii}(t+1).x_{ji}(t+1)} = P_2.\overline{x_{ii}(t).x_{ji}(t)} +\frac{2}{N_c} \left( \overline{F_{ij}(t)G_{ji}(t)} + \sum_{p \notin \{i,j\}} \overline{F_{ip}(t)x_{ji}(t)} \right) + \frac{2k}{N_T} \sum_{p \notin \{j, i\}} \overline{J_{jpi}(t)x_{ii}(t)} ,
\end{dmath}
with:
\begin{align}
    P_2 = 1 - \frac{2}{N_a} -\frac{2k}{N_c}.
\end{align}
Moreover, for $p \neq i$ and $j \neq i$ and $p \neq j$:
\begin{dmath}
     \overline{x_{ji}(t+1).x_{pi}(t+1)} =  P_3. \overline{x_{ji}(t).x_{pi}(t)} +\frac{2}{N_c} \left( \overline{x_{pi}(t)G_{ji}(t)} +  \overline{x_{ji}(t)G_{pi}(t)}\right) + \frac{2k}{N_T}\overline{J_{jpi}(t)J_{pji}(t)} + \frac{2k}{N_T} \sum_{q \notin \{j, i, p\}} \left(\overline{J_{jqi}(t)x_{pi}(t)} + \overline{J_{pqi}(t)x_{ji}(t)} \right),
\end{dmath}
with:
\begin{align}
    P_3 = 1- \frac{4}{N_c}-\frac{2k}{N_T} -\frac{4k(N_a-3)}{N_t}.
\end{align}

\end{document}